\documentclass{svjour2}                    
\smartqed  
\usepackage{graphicx}
\usepackage{latexsym}
\journalname{J. Stat. Phys.}

\newcommand{\be}{\begin{equation}}
\newcommand{\ee}{\end{equation}}
\newcommand{\af}{{ \lambda }} 
\newcommand{\qhat}{ {\hat Q} } 
\newcommand{\mfont}{ \mathcal } 
\newcommand{\siga}{{ \Sigma_{11} }} 
\newcommand{\sigb}{{ \Sigma_{12} }} 
\newcommand{\sigc}{{ \Sigma_{21} }} 
\newcommand{\sigd}{{ \Sigma_{22} }} 
\newcommand{\topsig}{{ \Sigma_T }}
\newcommand{\botsig}{{ \Sigma_B }}
\newcommand{\amp}{{ A_\phi }} 
\newcommand{\ampz}{{ a_\phi }} 
\newcommand{\period}{{ \Delta \tau }} 
\newcommand{\rat}{{ {\cal R} }}

\begin{document} 

\title{Hill's Equation with Random Forcing Parameters: 
Determination of Growth Rates through Random Matrices} 

\titlerunning{Hill's Equation with Random Forcing} 

\author{Fred C. Adams \and Anthony M. Bloch} 

\institute{F. C. Adams \at
Michigan Center for Theoretical Physics \\
Physics Department, University of Michigan, Ann Arbor, MI 48109\\
\email{fca@umich.edu}           
\and
A. M. Bloch \at
Michigan Center for Theoretical Physics \\
Mathematics Department, University of Michigan, Ann Arbor, MI 48109\\
\email{abloch@umich.edu}           
}

\date{Received: date / Accepted: date}
\maketitle

\begin{abstract} 

This paper derives expressions for the growth rates for the random 
$2 \times 2$ matrices that result from solutions to the random Hill's
equation. The parameters that appear in Hill's equation include the
forcing strength $q_k$ and oscillation frequency $\af_k$.  The
development of the solutions to this periodic differential equation
can be described by a discrete map, where the matrix elements are
given by the principal solutions for each cycle. Variations in the
$(q_k,\af_k)$ lead to matrix elements that vary from cycle to
cycle. This paper presents an analysis of the growth rates including
cases where all of the cycles are highly unstable, where some cycles
are near the stability border, and where the map would be stable in
the absence of fluctuations. For all of these regimes, we provide
expressions for the growth rates of the matrices that describe the
solutions.

\end{abstract} 


\section{Introduction} 

This paper considers the growth rates for Hill's equation with
parameters that vary from cycle to cycle. In this context, Hill's
equation takes the form
\be 
{d^2 y \over dt^2} + [ \af_k + q_k \qhat (t) ] y = 0 \, , 
\label{basic} 
\ee 
where the barrier shape function $\qhat(t)$ is periodic, so that
$\qhat (t + \period) = \qhat(t)$, where $\period$ is the period. Here
we take $\period = \pi$, and the function $\qhat$ is normalized so
that $\int_0^\period \qhat dt$ = 1. The forcing strength parameters
$q_k$ are a set of independent identically distributed (i.i.d.) random
variables that take on a new value every cycle (where the index $k$
labels the cycle). The parameters $\af_k$, which determine the
oscillation frequency in the absence of forcing, also vary from cycle
to cycle (and are i.i.d.). In principal, the cycle interval $\period$
could also vary; however, this generalized case can be reduced to the
problem of equation (\ref{basic}) through an appropriate re-scaling of
the other parameters (see Theorem 1 of [AB]).

Hill's equations [HI] with constant values of the parameters have been
well studied and arise in a wide variety of applications [MW]. The
introduction of parameters that sample a distribution of values is
thus a natural generalization of this classic problem. Here we refer
to the case with constant parameters as the ``classical regime'' of
the general case.

For this class of periodic differential equations, the transformation
that maps the coefficients of the principal solutions from one cycle
to the next takes the form
\be
{\mfont M}_k = \left[ 
\matrix{h_k & (h_k^2 - 1)/g_k \cr g_k & h_k} \right] \, , 
\label{mapzero} 
\ee
where the subscript denotes the cycle. The matrix elements are defined
by $h_k = y_1 (\pi)$ and $g_k = {\dot y}_1 (\pi)$ for the $kth$ cycle,
where $y_1$ and $y_2$ are the principal solutions for that cycle. Note
that the matrix has only two independent elements rather than four:
Since the Wronskian of the original differential equation
(\ref{basic}) is unity, the determinant of the matrix map must be
unity, and this constraint eliminates one of the independent elements.
In addition, this paper specializes to the case where the periodic
functions $\qhat(t)$ are symmetric about the midpoint of the period,
so that $y_1(\pi) = {\dot y}_2 (\pi)$, which eliminates a second
independent element [MW]; this symmetry applies to the applications 
that motivated this work.

For transformation matrices ${\mfont M}_k$ of the form
(\ref{mapzero}), the eigenvalues $\lambda_k$ can be used to classify
the matrix types [LR].  The characteristic polynomial has the form
\be
\lambda_k^2 - 2 h_k \lambda_k + 1 = 0 \, . 
\ee
This equation allows for three classes of eigenvalues $\lambda_k$: For
$|h_k| > 1$, the eigenvalues are real and have the same sign, and the
transformation matrix is hyperbolic symplectic; we denote this regime
as classically unstable.  When $|h_k| < 1$, the eigenvalues are
complex and the matrix is elliptic; this regime is denoted as
classically stable.  The remaining possibility is for $|h_k| = 1$,
which leads to degenerate eigenvalues equal to either $+1$ or $-1$;
these matrices are parabolic and are stable under multiplication.

This paper studies the multiplication of infinite strings of random
matrices of the form (\ref{mapzero}), i.e., the product of $N$ such
matrices in the limit $N \rightarrow \infty$.  The problem of finding
growth rates for infinite products of matrices with random elements
was formulated over four decades ago [FU, FK], where existence results
were given.  We recall the key result here for convenience: 

\noindent
For a $k\times k$ matrix $A$ with real or complex entries, let $||A||$
denote the Frobenius norm.

\noindent
{\bf Theorem (FK):} 
Let $X^1, X^2, X^3,\dots$ form a metrically transitive stationary
stochastic process with values in the set of $k\times k$ matrices. 
Suppose $\log^+||X^1||$ exists, where ${\log}^+ t=\max (\log t,0)$, 
then the limit $\lim_{N\rightarrow\infty}||X^NX^{N-1}\cdots X^1||$ exists.

Determination of the growth rates are thus carried out in the limit of
large $N$, and all probabilistic limits given here are meant almost
surely.

A great deal of subsequent work has studied differential equations of
the form (\ref{basic}) and the growth rates of the corresponding
random matrices [CL, PF, LGP]. In spite of this progress, there are
relatively few examples that provide explicit expressions for the
growth rates. The goal of this paper is relatively modest: It provides
(what we believe to be) new analytic expressions for the growth rates
of random matrices of the form (\ref{mapzero}). These expressions are 
derived for various regimes of parameter space, as described below.

The outline of this paper is as follows: Section 2 reviews the
astrophysical background that led us to this topic.  Section 3
considers matrix multiplication for the case where the solutions are
unstable in the classical regime.  Section 4 develops approximations
for this regime and provides some numerical verification. Section 5
considers matrix multiplication in the regime where the solutions are
classically stable. In this case, the transformation matrices
$\mfont{M}_k$ correspond to elliptical rotations and matrix
multiplication is stable in the absence of fluctuations; random
variations in the matrix elements render the solutions unstable. The
paper concludes (in Section 6) with a brief summary of the results.

\section{Astrophysical Background}
 
The motivation for considering random Hill's equations arose in
studies of orbit problems in astrophysics [AK].  When an orbit starts
in the principal plane of a triaxial, extended mass distribution (such
as a dark matter halo), the motion is unstable to perturbations in the
perpendicular direction. The development of the instability is
described by a random Hill's equation with the form given by equation
(\ref{basic}).

To illustrate this type of behavior, consider an extended mass
distribution with a density profile of the form
\be
\rho = {\rho_0 \over m} \qquad {\rm with} \qquad 
m^2 = {x^2 \over a^2} + {y^2 \over b^2} + {z^2 \over c^2}  ,
\label{halodef} 
\ee
where $\rho_0$ is a density scale. This form arises in many different
astrophysical contexts, including dark matter halos, galactic bulges,
and young embedded star clusters.  The density field is thus constant
on ellipsoids, where, without loss of generality, $a > b > c > 0$.
For this density profile, one can find analytic forms for both the
gravitational potential and the force terms [AK]. From these results,
one can determine the orbital motion for a test particle moving in the
potential resulting from the triaxial density distribution of equation
(\ref{halodef}).  When the orbit begins in any of the three principal
planes, the motion is generally unstable to perturbations in the
perpendicular direction [AB, AK]. For example, for an orbit initially
confined to the $x$-$z$ plane, the amplitude of the $y$ coordinate
will (usually) grow exponentially with time. In the limit of small
$|y| \ll 1$, the equation of motion for the perpendicular coordinate
simplifies to the form
\be
{d^2 y \over dt^2} + \omega_y^2 y = 0 
\qquad {\rm where} \qquad \omega_y^2 = 
{ 4/b \over \sqrt{c^2 x^2 + a^2 z^2} + b \sqrt{x^2 + z^2} } \ . 
\label{omegay}
\ee
The time evolution of the coordinates $(x,z)$ is determined by the
orbit in the original $x$-$z$ plane. Since the orbital motion is
nearly periodic, the $[x(t),z(t)]$ dependence of $\omega_y^2$
represents a nearly periodic forcing term. The forcing strengths, and
hence the parameters $q_k$ appearing in Hill's equation (\ref{basic}), 
are determined by the inner turning points of the orbit (with appropriate 
weighting from the axis parameters $[a,b,c]$). Since the orbits are
usually chaotic, the distance of closest approach, and hence the
strength $q_k$ of the forcing, varies from cycle to cycle. The outer
turning points of the orbit provide a minimum value of $\omega_y^2$,
which defines the unforced oscillation frequency $\af_k$ appearing in
Hill's equation. As a result, the quantity $\omega_y^2$ can be written 
in the form 
\be
\omega_y^2 = \af_k + Q_k (t) \, ,
\ee 
where the index $k$ counts the number of orbit crossings. The shapes
of the functions $Q_k$ are nearly the same, so that one can write
$Q_k$ = $q_k {\hat Q}(t)$, where ${\hat Q}(t)$ is periodic. The
chaotic orbit in the original plane leads to different values of
$\af_k$ and $q_k$ for each crossing. The equation of motion
(\ref{omegay}) for the $y$ coordinate thus takes the form of Hill's
equation (\ref{basic}), where the period, forcing strength, and
oscillation frequency vary from cycle to cycle.

\section{Matrix Multiplication for the Classically Unstable Regime} 

The goal of this work is to find growth rates for solutions of the
differential equation (\ref{basic}). These growth rates are determined
by multiplication of the random matrices ${\mfont M}_k$ (from equation
[\ref{mapzero}]) that connect solutions from cycle to cycle.  These
transformation matrices can also be written in the form
\be 
{\mfont M}_k = h_k {\mfont B}_k \qquad {\rm where} \qquad 
{\mfont B}_k = \left[ \matrix{1 & x_k \phi_k \cr {1/x_k} & 1} \right] \, , 
\label{mbdefine} 
\ee
where $x_k$ = $h_k/g_k$ and $\phi_k$ = $1 - 1/h_k^2$. By virtue of 
our assumption on the variables ($q_k$, $\af_k$), the matrices 
${\mfont M}_k$ form a sequence of i.i.d. matrices.  In this section,
we consider the problem of matrix multiplication with matrices of the
form (\ref{mbdefine}). We specialize to the case where the solutions
are unstable in the classical regime so that $|h_k| \ge 1$ and to the
case where $x_k > 0$.  We also assume that the $h_k$, $x_k$, and
$1/x_k$ have finite means.  With the matrices written in the form
(\ref{mbdefine}), the highly unstable regime considered in [AB] can be
defined as follows:

\noindent
{\bf Definition:} Given that solutions to Hill's equation
(\ref{basic}) are determined by transformation matrices of the form
(\ref{mbdefine}), the {\it highly unstable regime} is defined by
setting $\phi_k = 1$. This specification thus defines a restricted problem. 

We remark that the above regime applies when the matrix elements
$|h_k| \gg 1$, which occurs for forcing strength parameters 
$q_k \gg 1$ [AB2]. 

The growth rates for Hill's equation (\ref{basic}) are determined 
by the growth rates for matrix multiplication of the full set of 
matrices ${\mfont M}_k$.  For a given matrix product, denoted here 
as ${\mfont M}^{(N)}$, the {\it growth rate} $\gamma$ is determined by 
\be
\gamma = \lim_{N \to \infty} {1 \over N} 
\log || {\mfont M}^{(N)} || \, , 
\label{growbasic}
\ee
where the result is independent of the choice of norm $|| \cdot ||$.
We note that the growth rate is called the {\it top} or 
{\it largest Lyapunov exponent}.

Equation (\ref{mbdefine}) separates the growth rate for this problem
into two parts. Let the expectation value of a sequence $X_k$
be denoted by 
$$\langle X_k\rangle=\lim_{N \to \infty}
{1 \over N} \sum_{k=1}^N X_k$$
Then the first part $\gamma_h$ of the growth rate is given
by
\be
\gamma_h = \lim_{N \to \infty} {1 \over N} \sum_{k=1}^N \log |h_k| 
=\langle\log |h_k|\rangle \, .
\ee 
We limit our discussion to distributions of the $h_k$ for which this
limit is finite.  The remaining part of the growth rate is determined
by matrix multiplication of the ${\mfont B}_k$.  Note that the
original differential equation (\ref{basic}) is defined on a time
interval $0 \le t \le \pi$, so that the definition of its growth rate
includes a factor of $\pi$ [MW], whereas the growth rate for matrix
multiplication (\ref{growbasic}) generally does not [FK]. Ignoring
these normalization issues, this paper focuses on the calculation of
the growth rates for the matrices ${\mfont M}_k$ and ${\mfont B}_k$.

The product of $N$ matrices of type ${\mfont B}_k$ 
can be written in the form 
\be
{\mfont B}^{(N)} \equiv 
\prod_{k=1}^N {\mfont B}_k = \left[ \matrix{\siga & x_1 \sigb \cr 
(1 / x_1) \sigc & \sigd } \right] \, ,  
\label{product} 
\ee 
where the first equality defines notation and where 
$$
\siga = \sum_{j=1}^{2^{N-1}} r_j a_j \, , \qquad 
\sigb = \sum_{j=1}^{2^{N-1}} r_j b_j \, , 
$$
\be 
\sigc = \sum_{j=1}^{2^{N-1}} {1 \over r_j} c_j \, , \qquad 
\sigd = \sum_{j=1}^{2^{N-1}} {1 \over r_j} d_j \, . 
\ee
Here, the variables $r_j$ are products of ratios of the form 
\be
r_j = {x_{\mu_1} x_{\mu_2} \dots x_{\mu_n} \over 
x_{\nu_1} x_{\nu_2} \dots x_{\nu_n} } \, .
\ee
The indices are confined to the range $1 \le \mu_i, \nu_i \le N$. 
The additional factors $a_j$, $b_j$, $c_j$, $d_j$ are products of 
the variables $\phi_j$, and can be written in the form
\be
a_j = \prod_{k=1}^N \phi_k^{p_k} \qquad {\rm where} \qquad 
p_k = 0 \, \, \, {\rm or} \, \, \, 1 \, \, . 
\ee 

\medskip 
\noindent 
{\bf Result 1:} For the case where $|h_k| > 1$ for all cycles,
and in the limit of large $N$, the eigenvalue of the product matrix
is given by the formula 
\be
\lambda = \siga + \sigd \, + {\cal O} \left( h^{-2N} \right) \, ,  
\label{result1} 
\ee
where each of these quantities should be labeled at the 
$Nth$ iteration. 

\noindent 
{\it Proof:} The characteristic equation of the product matrix 
of equation (\ref{product}) takes the form 
\be 
\lambda^2 - \lambda (\siga + \sigd) + \siga \sigd - \sigb \sigc = 0 \, . 
\label{charact} 
\ee
The final term is the determinant of the product matrix, and this
determinant is given by the product of the individual matrices, so
that
\be
\siga \sigd - \sigb \sigc = \prod_{k=1}^N (1 - \phi_k) = 
\prod_{k=1}^N {1 \over h_k^2} \, . 
\ee 
Given that $|h_k| > 1$ $\forall k$, this term vanishes in the limit 
$N \to \infty$.  As a result, the growing eigenvalue of the 
characteristic equation (\ref{charact}) simplifies to the form
$\lambda = \siga + \sigd$. $\Box$

\medskip 
\noindent
{\bf Result 2:} The four sums that specify the matrix elements of
the product matrix are not independent. In particular, for the case
where $|h_k| > 1$ and in the limit $N \to \infty$, the ratios of the
matrix elements approach the form
\be
{\sigb  \over \siga} = {\sigd \over \sigc} = {\rm constant} \equiv f \, . 
\label{result2} 
\ee 

\noindent
{\it Proof:} As shown above, the determinant of the product matrix vanishes 
in the limit $N \to \infty$, so that in the limit
\be
\siga \sigd = \sigb \sigc \, . 
\ee
The result implied by the first equality of equation (\ref{result2}) 
follows immediately. 

Further, one can show by direct construction that if the relation of
equation (\ref{result2}) holds, then the relation is preserved under
matrix multiplication. Let the product matrix after $N$ cycles have
the form
\be
{\mfont B}^{(N)} = \left[ \matrix{ \topsig & f x_1 \topsig \cr 
(1/x_1) \botsig & f \botsig} \right] \, , 
\label{ncycle} 
\ee
where $f$ is the constant in equation (\ref{result2}). Then the
matrix takes the following from after the next cycle:
\be
{\mfont B}^{(N+1)} = 
\left[ \matrix{ \topsig + (x/x_1) \phi \botsig & x_1 
f ( \topsig + (x/x_1) \phi \botsig ) \cr 
(1/x_1) (\botsig + (x_1/x) \topsig ) & 
f (\botsig + (x_1/x) \topsig) } \right] \, , 
\label{andone} 
\ee
so that the left-right symmetry relation is conserved. $\Box$ 

In the above proof we have adopted notation that is used throughout
this paper: The subscript `1' denotes the values of the parameters
(e.g., $x_1$) for the first cycle in the series. Since the results of
this problem can be written in terms of this starting value, these
initial values play a recurring role. The subscript `$N$' denotes the
values of the parameters (e.g, $x_N$) appropriate for the $N$th cycle
of the series. In iteration formulae, however, we use unsubscripted
variables (e.g., $x$) for the next ($N+1$)st cycle.

\medskip 
\noindent
{\bf Result 3:} In the highly unstable regime, the ratio of 
$\topsig$ to $\botsig$ has the form:
\be 
{\topsig \over \botsig} = {x \over x_1} \, . 
\label{result3} 
\ee 

\noindent
{\it Proof:} From our previous results (see equation [19] of [AB]),
the product matrix after $N$ cycles has the form given by equation
(\ref{ncycle}) with $f = 1$ (in the highly unstable regime). After one
additional multiplication, we obtain the form given by equation
(\ref{andone}) with $f$ = 1. We thus find
\be
{\topsig^{(N+1)} \over \botsig^{(N+1)} } = 
{\topsig^{(N)} + (x/x_1) \botsig^{(N)} \over \botsig^{(N)} + 
(x_1/x) \topsig^{(N)} } = {x \over x_1} \,  . 
\ee
For each cycle the ratio $x/x_1$ has a different value, so that no 
limit is reached as $N \to \infty$. However, the ratio at any given
finite cycle obeys equation (\ref{result3}). $\Box$ 

To derive an expression for the growth rate for matrix
multiplication, we first define
\be 
S \equiv \siga + \sigd \, . 
\ee 
As shown in the proof of Result 1, the eigenvalue of the
product matrix approaches $S$, as defined above, in the limit 
$N \to \infty$.  By construction, the iteration formula for $S$ 
takes the form 
\be
S^{(N+1)} = S^{(N)} \left[ 1 + 
{ (x/x_1) \phi \sigc^{(N)} + (x_1/x) \sigb^{(N)} \over 
\siga^{(N)} + \sigd^{(N)} } \right] \, . 
\ee
Using the definition of $f$, $\topsig$, and $\botsig$, this expression 
can be simplified to the form 
\be
S^{(N+1)} = S^{(N)} \left[ 1 + 
{ (x/x_1) \phi \botsig^{(N)} + (x_1/x) f \topsig^{(N)} \over 
\topsig^{(N)} + f \botsig^{(N)} } \right] \, . 
\label{siterate} 
\ee

\medskip 
\noindent
{\bf Result 4:} In the highly unstable regime the
iteration formula for the eigenvalue reduces to the form 
\be
S^{(N+1)} = S^{(N)} \left[ 1 + {x_N \over x} \right] \, . 
\label{result4}
\ee
This result agrees with that of Theorem 2 from [AB]. 

\noindent 
{\it Proof:} In the highly unstable regime $\phi = 1$, $f = 1$, and 
equation (\ref{result3}) holds for the ratio of $\topsig/\botsig$. 
The iteration formula of equation (\ref{siterate}) thus reduces to 
\be
S^{(N+1)} = S^{(N)} \left[ 1 + 
{ (x/x_1) + (x_N/x) \over 1 + {x_N/x_1} } \right] 
= S^{(N)} \left[ 1 + {x_N \over x} \right] 
\left[ { x_1 + x \over x_1 + x_N } \right] \, . 
\label{iteratehu} 
\ee
Since the starting value $x_1$ is fixed, the second factor in 
square brackets approaches unity in the limit $N \to \infty$, i.e., 
\be
\lim_{N \to \infty} \prod_{k=1}^N 
\left[ { x_1 + x_{k+1} \over x_1 + x_k } \right] = 1 \, . 
\ee 
The expression of equation (\ref{iteratehu}) thus reduces to
that of equation (\ref{result4}). $\Box$

Motivated by the result of equation (\ref{result3}) for the highly
unstable regime, we write the ratio of matrix elements for the general
case in the form
\be
{\topsig^{(N)} \over \botsig^{(N)}} = {x_N \over x_1} \alpha_N \, , 
\label{alphadef}
\ee
so that 
\be
S^{(N+1)} = S^{(N)} \left[ 1 + 
{ (x/x_1) \phi + (x_N/x) f \alpha_N \over f + \alpha_N (x_N/x_1) } 
\right] \equiv {\cal F}_N S^{(N)} \, ,
\label{sitfactor} 
\ee
where the second equality defines ${\cal F}_N$.  The parameter
$\alpha_N$ incorporates the correction due to the matrices not being
in the highly unstable regime. Note that $f$ approaches a constant
value (from Result 2) and $x_1$ is a constant (by definition).  The
iteration factor ${\cal F}_N$ can be rewritten in the form
\be
{\cal F}_N = \left[ 1 + 
{ x^2 \phi + b \alpha_N x_N \over x (b + \alpha_N x_N) } \right] 
\qquad {\rm where} \qquad b \equiv f x_1 \, . 
\label{itfactor} 
\ee

\noindent
{\bf Theorem 1:} The growth rate for matrix multiplication, with
products of the general form defined through equation (\ref{product}),  
is given by
\be 
\gamma = \lim_{N \to \infty} {1 \over N} \sum_{k=1}^N \log \left[ 1 + 
{ x_k^2 \phi_k + \alpha_{k-1} x_{k-1} \over x_k (1 + \alpha_{k-1} x_{k-1}) } 
\right] \, , 
\label{gammafull} 
\ee 
where the $\alpha_k$ are determined through the iteration formula 
\be 
\alpha_{k} = {x_k \phi_k + x_{k-1} \alpha_{k-1} \over 
x_k + x_{k-1} \alpha_{k-1}} \, .  
\label{iteralpha} 
\ee 

\noindent 
{\it Proof:} Note that existence of the required limit holds by the
Theorem of FK. Equations (\ref{sitfactor} -- \ref{itfactor}) show 
that the growth rate is given by 
\be
\gamma = \lim_{N \to \infty} {1 \over N} \sum_{k=1}^N \log {\cal F}_k = 
\lim_{N \to \infty} {1 \over N} \sum_{k=1}^N \log \left[ 1 + 
{ x_k^2 \phi_k + b \alpha_{k-1} x_{k-1} \over x_k (b + \alpha_{k-1} x_{k-1}) } 
\right] \, , 
\ee
where this form is exact, provided that the $\alpha_k$ are properly
specified. This issue is addressed below. To complete the proof, we
must also show that the growth rate is independent of the value of $b$,
so that we can set $b=1$ in the above formula. The derivative of the 
growth rate with respect to the parameter $b$ takes the form 
\be
{d \gamma \over db} = \lim_{N \to \infty} {1 \over N} 
\sum_{k=1}^N {1 \over {\cal F}_k} {d {\cal F}_k \over db} \, ,
\ee
which can be evaluated to take the form 
\be 
{d \gamma \over db} = \lim_{N \to \infty} {1 \over N} \sum_{k=1}^N 
{ (\alpha_{k-1} x_{k-1})^2 - x_k^2 \phi_k \over (b + \alpha_{k-1} x_{k-1}) 
\left[ x_k (b + \alpha_{k-1} x_{k-1}) + x_k^2 \phi_k + 
b \alpha_{k-1} x_{k-1} \right] } \, . 
\ee
This expression vanishes in the limit.

To show that the $\alpha_k$ are given by equation (\ref{iteralpha}),
we start with the result of matrix multiplication from equation
(\ref{andone}) and use the definition of $\alpha_k$ from equation
(\ref{alphadef}); these two results imply that
\be
\alpha_{k+1} = {x_1 \over x_{k+1}} { \topsig^{(k+1)} \over \botsig^{(k+1)} }
= {x_1 \over x_{k+1}} { \topsig^{(k)} + (x_{k+1}/x_1) \phi_{k+1} \botsig^{(k)} 
\over \botsig^{(k)} + (x_1/x_{k+1}) \topsig^{(k)} } \, .  
\ee 
We can then eliminate the factors of $\topsig$ and $\botsig$ by again 
using the definition of $\alpha_k$ from equation (\ref{alphadef}), 
and thus obtain 
\be 
\alpha_{k+1} = {x_1 \over x_{k+1}} { (x_k/x_1) \alpha_k + (x_{k+1}/x_1)
\phi_{k+1} \over 1 + (x_k/x_{k+1}) \alpha_k } \, = \, 
{ x_k \alpha_k + x_{k+1} \phi_{k+1} \over x_{k+1} + x_k \alpha_k} \, .  
\ee 
After re-labeling the indices, we obtain equation (\ref{iteralpha}). 
$\Box$

\section{Approximations for the Classically Unstable Regime} 

For classically unstable matrices with $|h_k| > 1$, Theorem 1 provides
an exact expression for the growth rate. Since the formulae are
complicated, this section presents simpler but approximate expressions
for the growth rates for the case where $\phi_k$ are small (Theorem 2)
and where the differences $1 - \phi_k$ are small (Theorem 3). We also
present two heuristic approximations for the growth rates for the
general problem.

\medskip
\noindent 
{\bf Theorem 2:} In the regime where the variables $\phi_k$ are small,
$\phi_k x_k \ll 1$ $\forall k$, the growth rate for the matrix 
${\mfont B}_k$ tends in the limit of large $N$ to the form:
\be
\gamma = \log \left( 1 + \left[ \langle {1 / x_k} \rangle 
\langle {x_k \phi_k} \rangle \right]^{1/2} \right)  + 
{\cal O} \left( \langle {x_k \phi_k} \rangle \right) \, . 
\label{theorem2} 
\ee

\noindent 
{\it Proof:} We first break up the matrix into two parts so that 
${\mfont B}_k$ = ${\mfont I}$ + ${\mfont A}_k$, 
where $\mfont{I}$ is the identity matrix and where 
\be
{\mfont A}_k = \left[ \matrix{0 & x_k \phi_k \cr 1/x_k & 0} \right] = 
\left[ \matrix{0 & \eta_k \cr y_k & 0} \right] \, . 
\label{defineak} 
\ee
Note that the second equality defines $\eta_k = x_k \phi_k$ and 
$y_k = 1/x_k$.  We first show (by induction) that repeated
multiplications of the matrices ${\mfont A}_k$ lead to products 
with simple forms. The products of even numbers $N = 2 \ell$ of 
matrices $\mfont{A}_k$ produce diagonal matrices of the form 
\be
{\mfont A}^{(N)} = {\mfont A}^{(2 \ell)} = \prod_{k=1}^N {\mfont A}_k = 
\left[ \matrix{P_\ell^A & 0 \cr 0 & P_\ell^B} \right] \, , 
\label{aproducteven} 
\ee
where the products $P_\ell$ are defined by 
\be
P_\ell^A = \prod_{i=1}^\ell 
\left( \eta_{2i} \right) \left( y_{2i-1} \right)  
\qquad {\rm and} \qquad 
P_\ell^B = \prod_{i=1}^\ell 
\left( \eta_{2i-1} \right) \left( y_{2i} \right) \, . 
\label{pells} 
\ee
Similarly, the product of odd numbers $N = 2 \ell + 1$ of matrices 
$\mfont{A}_k$ produce off-diagonal matrices of the form 
\be
{\mfont A}^{(N)} = {\mfont A}^{(2 \ell+1)} = \prod_{k=1}^N {\mfont A}_k = 
\left[ \matrix{0 & Q_\ell^B \eta_1 \cr Q_\ell^A y_1 & 0 } \right] \, , 
\label{aproductodd} 
\ee
where the products $Q_\ell$ are defined analogously to the $P_\ell$. 
The product of $N$ matrices $\mfont{B}_k$ can then be written in the form 
\be
\mfont{B}^{(N)} = \prod_{k=1}^N \mfont{B}_k = \left[ 
\matrix{\siga & \sigb \eta_1 \cr \sigc y_1 & \sigd } \right] \, . 
\ee
Without loss of generality, let $N = 2 \ell$ be even. 
Then the matrix elements are given by 
$$ 
\siga = \sum_{\ell=0}^{N/2} \sum_{j=1}^{C^N_{2\ell}} 
\left( P_\ell^A \right)_j \, , \qquad 
\sigd = \sum_{\ell=0}^{N/2} \sum_{j=1}^{C^N_{2\ell}} 
\left( P_\ell^B \right)_j \, , \qquad 
$$
\be
\sigb = \sum_{\ell=0}^{N/2-1} \sum_{j=1}^{C^N_{2\ell+1}} 
\left( Q_\ell^B \right)_j \, , \qquad 
\sigc = \sum_{\ell=0}^{N/2-1} \sum_{j=1}^{C^N_{2\ell+1}} 
\left( Q_\ell^A \right)_j \, , \qquad 
\ee
where $C^N_\ell$ is the binomial coefficient and where the subscripts
on the $P_\ell$ and $Q_\ell$ denote different realizations of the
products.

The eigenvalue $\Lambda_N$ of the product matrix at the $Nth$ iteration
is given by its characteristic equation, which has the solution 
\be
\Lambda_N = {1 \over 2} \left\{ \siga + \sigd + \left[ 
(\siga - \sigd)^2 + 4 \sigb \sigc \eta_1 y_1 \right]^{1/2} \right\} \, . 
\ee 
In the limit of large $N$, we can make the approximation that 
$\siga \approx \sigd$ and $\sigb \approx \sigc$, so that the
expression for the eigenvalue takes the form
\be
\Lambda_N = \siga + \sigb \left[\eta_1 y_1 \right]^{1/2} 
= \sum_{\ell=0}^{N/2} \sum_{j=1}^{C^N_{2\ell}} 
\left( P_\ell^A \right)_j + 
\sum_{\ell=0}^{N/2-1} \sum_{j=1}^{C^N_{2\ell+1}} 
\left( Q_\ell^B \right)_j \left[\eta_1 y_1 \right]^{1/2} .
\ee
In the limit of large $N$, all the binomial coefficients are 
large except for the first and last one. We can thus rewrite the 
above equation in the form 
\be
\Lambda_N = \sum_{\ell=0}^{N/2} C^N_{2\ell} 
\left( \left\langle P_\ell^A \right\rangle + \varepsilon_\ell \right) 
+ \sum_{\ell=0}^{N/2-1} C^N_{2\ell+1} 
\left( \left\langle Q_\ell^B \right\rangle + \varepsilon_\ell \right) 
\left[\eta_1 y_1 \right]^{1/2} \, . 
\ee
If the realizations of the products $(P_\ell)_j$ were independent, the
error terms $\varepsilon_\ell$ would vanish in the limit. However, for a 
given $N$, the sums contain $C_{2\ell}^N$ terms, and $C_{2\ell}^N > N$ 
in general, so all of the terms in the sum cannot be independent. 
We then write the products $\left\langle P_\ell^A \right\rangle$ and
$\left\langle Q_\ell^B \right\rangle$ in the form
\be
\left\langle P_\ell^A \right\rangle + \varepsilon_\ell = 
\langle \eta_j \rangle^\ell \langle y_j \rangle^\ell 
(1 + \epsilon_\ell)^\ell \, ,   
\ee
and similarly for $\left\langle Q_\ell^B \right\rangle$. This form is
exact if one uses the proper expressions for the $\epsilon_\ell$. 
Using this result, the expression for the eigenvalue $\Lambda_N$ becomes
\be
\Lambda_N = \sum_{\ell=0}^{N/2} C^N_{2\ell} 
\langle \eta_j \rangle^\ell \langle y_j \rangle^\ell (1 + \epsilon_\ell)^\ell
+ \sum_{\ell=0}^{N/2-1} C^N_{2\ell+1} 
\langle \eta_j \rangle^\ell \langle y_j \rangle^\ell (1 + \epsilon_\ell)^\ell
\left[\eta_1 y_1 \right]^{1/2} \, , 
\ee
which takes the form 
\be
\Lambda_N = \sum_{k=0}^{N} C^N_k
\langle \eta_j \rangle^{k/2} \langle y_j \rangle^{k/2} 
(1 + \epsilon_{k} )^{k/2} \, . 
\ee
If we expand this result, we find that 
\be
\Lambda_N = 1 + N \langle \eta_j \rangle^{1/2} \langle y_j \rangle^{1/2} 
(1 + \epsilon_{1} )^{1/2} + C_2^N \langle \eta_j \rangle \langle y_j \rangle
(1 + \epsilon_{2} ) + \dots 
\ee 
Further, by performing an exact treatment of the first order expansion 
[AB2] we find that $\epsilon_1$ = 0. This finding allows us to write the  
product in the form 
\be
\Lambda_N = \left[ 1 + \langle \eta_j \rangle^{1/2} \langle y_j \rangle^{1/2} 
+ {\cal O} (\eta_j) \right]^N \, . 
\ee
The growth rate thus becomes 
\be 
\gamma = \log \left[ 1 + \langle \eta_j \rangle^{1/2}
\langle y_j \rangle^{1/2} \right] + {\cal O} (\eta_j) \, . 
\ee
This last expression is valid provided that $\eta_j \ll 1$ $\forall j$. 
$\Box$ 

Note that to consistent order, we can replace the limiting form of
equation (\ref{theorem2}) with the equivalent, simpler function
\be
\gamma \to 
\left[ \langle 1/x_k \rangle \langle \eta_k \rangle \right]^{1/2} \, . 
\ee 

\begin{figure} 
\centering
\includegraphics[width=120mm]{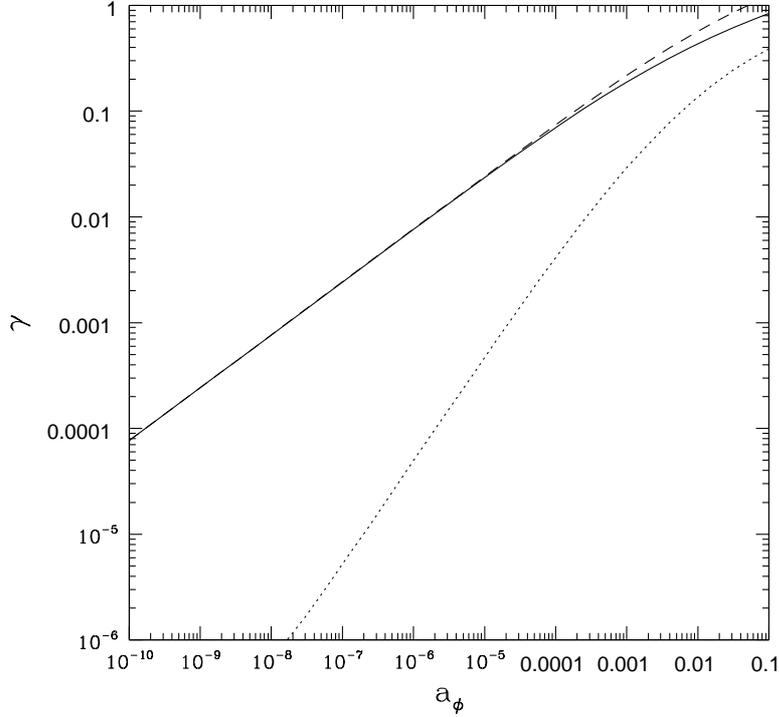} 
\caption{Growth rates for small $\phi_k$. The variables $\phi_k$ are
determined through the relation $\phi_k = \ampz \, \xi_k$, where
$\xi_k$ is uniformly distributed on [0,1].  The solid curve shows the
growth rate $\gamma$ calculated directly from matrix multiplication as
a function of the amplitude $\ampz$. The dashed curve shows the
estimate $\gamma_2$ for the growth rate from Theorem 2. The dotted
curve shows the difference $\Delta \gamma$ = $\gamma_2 - \gamma$.
Note that $\gamma \propto \sqrt{\ampz}$ whereas 
$\Delta \gamma \propto \ampz$. } 
\label{fig:smallamp} 
\end{figure}

Figure \ref{fig:smallamp} illustrates how well the approximation of
Theorem 2 works. For the sake of definiteness, the variables $x_k$ 
are log-uniformly distributed with $\log_{10} x_k \in [-2,2]$.  The 
$\phi_k$ obey the relation $\phi_k = a_\phi \xi_k$, where $\xi_k$ is a
uniformly distributed random variable over the interval $[0,1]$.  As
shown by the figure, the limiting form of equation (\ref{theorem2}) 
provides an excellent description of the calculated growth rate for
sufficiently small $\phi_k$.

Next we consider the case where the correction factors $\phi_k$ are
close to unity. In this case the variables $(1-\phi_k) \ll 1$, and we
can expand to leading order in $(1-\phi_k)$. This procedure leads to 
the following result:

\medskip 
\noindent 
{\bf Theorem 3:} Let $\gamma_0$ be the growth rate for the highly
unstable regime where $\phi_k = 1$. For small perturbations about 
this limiting case, the growth rate takes the form $\gamma$ = 
$\gamma_0 - \delta \gamma$, where 
\be 
\delta \gamma = \lim_{N \to \infty} {1 \over N} \sum_{k=1}^N  
{(1 - \phi_k) x_k^2 \over (x_{k+1} + x_k) (x_k + x_{k-1}) } \, + 
{\cal O} \left( \langle x_k^2 (1 - \phi_k)^2 \rangle \right) \, . 
\label{theorem3} 
\ee 

\noindent 
{\it Proof:} We again break up the matrix into two parts, 
\be
{\mfont B}_k = {\mfont C}_k - \epsilon_k {\mfont Z} 
\qquad {\rm with} \qquad {\mfont Z} \equiv 
\left[ \matrix{0 & 1 \cr 0 & 0} \right] \, , 
\label{decomp} 
\ee
where here $\epsilon_k \equiv x_k (1 - \phi_k)$ and ${\mfont C}_k$ is 
the matrix appropriate for the highly unstable regime. Note that  
${\mfont Z}$ does not depend on the index $k$. Here we work to 
first order in the small parameter $\epsilon_k$. After $N$ cycles, 
the product matrix takes the form 
\be
{\mfont B}_k^{(N)} = \prod_{k=1}^N {\mfont B}_k = {\mfont C}_k^{(N)} - 
\sum_{k=1}^N \epsilon_k {\mfont P}_k^N \, + {\cal O}(\epsilon_k^2) \, , 
\label{productb}
\ee
where the partial product matrices ${\mfont P}_k^N$ are given by 
\be 
{\mfont P}_k^N = \left\{ \prod_{j=k+1}^N {\mfont C}_j \right\} \, {\mfont Z} 
\left\{ \prod_{j=1}^{k-1} {\mfont C}_j \right\} \, . 
\ee
We ignore the case where the ${\mfont Z}$ factors appear on the
ends -- this effect is ${\cal O} (1/N)$ and vanishes in the limit.
The products of the ${\mfont C}_k$ matrices can be written in the form 
\be
{\mfont C}_k^{(N)} = \Sigma_T^N \left[ \matrix{1 & x_1 \cr 1/x_N & x_1/x_N} 
\right] \qquad {\rm where} \qquad \Sigma_T^N = \prod_{j=2}^N 
\left(1 + {x_j \over x_{j-1} } \right) \, , 
\label{productc} 
\ee
where these results follow from previous work [AB]. As a result, the 
matrices ${\mfont P}_k^N$ can be evaluated: 
\be
{\mfont P}_k^N = {x_k \Sigma_T^N \over (x_k + x_{k+1}) (x_{k-1} + x_{k}) } 
\left[ \matrix{1 & x_{1} \cr 1/x_N & x_{1}/x_N} \right] = 
{x_k \over (x_k + x_{k+1}) (x_{k-1} + x_{k}) } {\mfont C}_k^{(N)} \, . 
\ee 
The product matrix ${\mfont B}_k^{(N)}$, given by equation (\ref{productb}) 
to leading order, can now be written in the form
\be
{\mfont B}_k^N = {\mfont C}_k^N \left[ 1 - \sum_{k=1}^N 
{(1 - \phi_k) x_k^2 \over (x_k + x_{k+1}) (x_{k-1} + x_{k}) } \right] \, . 
\ee
The first factor is the product of the matrices for the highly
unstable regime. Since the second factor is a function (not a matrix)
its contribution to the growth rate is independent of the first factor
and represents a correction to the growth rate of the form
\be
\delta \gamma = \lim_{N \to \infty} {1 \over N} \sum_{k=1}^N 
{(1 - \phi_k) x_k^2 \over (x_k + x_{k+1}) (x_{k-1} + x_{k}) } 
+ {\cal O} (\epsilon_k^2) \, , 
\ee
where the equalities hold to leading order.  This correction to the
growth rate has the form given by equation (\ref{theorem3}). $\Box$

\begin{figure} 
\centering
\includegraphics[width=120mm]{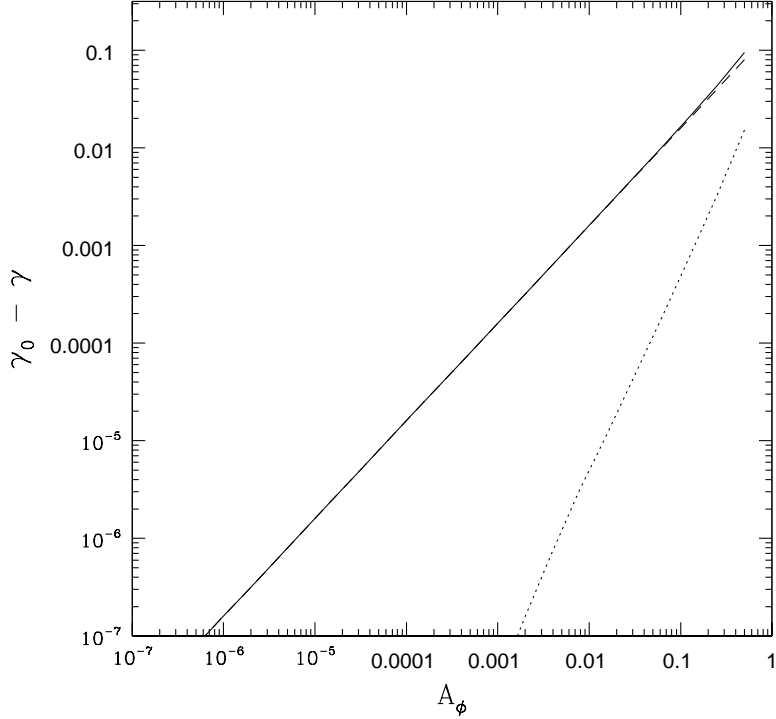} 
\caption{Growth rates for $\phi_k$ near unity. The variables $\phi_k$
are determined through the relation $\phi_k = 1 - \amp \, \xi_k$,
where $\xi_k$ is uniformly distributed on [0,1].  The solid curve
shows the quantity $\delta \gamma = \gamma_0 - \gamma$, where 
$\gamma$ is the growth rate calculated from matrix multiplication and
$\gamma_0$ is the growth rate for the highly unstable regime 
($\phi_k$ = 1 $\forall k$).  The dashed curve shows the estimate
$(\delta \gamma)_3 = (\gamma_0 - \gamma)_3$ for the difference in
growth rate calculated from Theorem 3.  The dotted curve shows the
error $\Delta$ = $(\delta \gamma)_3 - \delta \gamma$.  Note that
$\delta \gamma \propto \amp$ whereas the error term 
$\Delta \propto (\amp)^2$. }
\label{fig:largegam} 
\end{figure}

Figure \ref{fig:largegam} shows the growth rate for small departures
from the highly unstable regime. The correction factors are taken to
have the form $\phi_k = 1 - \amp \xi_k$, where $\xi_k$ is a uniformly
distributed random variable over the interval $[0,1]$. The highly
unstable regime corresponds to $\amp \to 0$.  The figure shows the
growth rate calculated from direct matrix multiplication (solid curve)
and the approximation from Theorem 3 (dashed curve) plotted as a
function of the amplitude $\amp$. Both curves plot the difference
$\gamma_0 - \gamma$, where $\gamma_0$ is the growth rate for the
highly unstable regime (where the $\phi_k$ = 1).

Since the general case is quite complicated it is useful to have
a good working approximation for the 
case where one is not in one of the two regimes
$\phi_k$ small or near unity.
 Toward this end, we first show that the values of $\alpha_k$
have a limited range:

\medskip 
\noindent
{\bf Result 5:} The variables $\alpha_k$ are confined to the range 
$\phi_{\rm min} \le \alpha_k \le 1$, where $\phi_{\rm min}$ is the 
minimum value of $\phi_k$. 

\noindent
{\it Proof:} We can rewrite the iteration formula (\ref{iteralpha}) 
for $\alpha_k$ in the alternate form 
\be
\alpha_{k} = {\phi_k + \beta_k \over 1 + \beta_k } \, , 
\ee
where we have defined the composite random variable  
$\beta_{k} \equiv \alpha_{k-1} x_{k-1} / x_k$. 
In the present context, $0 \le \beta_k < \infty$, and we can show that 
\be 
{d \alpha_k \over d \beta_k} > 0 
\ee
for all values of $\beta_k$. In the limit $\beta_k \to \infty$,
$\alpha_{k} \to 1$, whereas in the limit $\beta_k \to 0$,
$\alpha_{k} \to \phi$. Hence $\phi \le \alpha_k \le 1$ for all cycles. 
But $\phi \ge \phi_{\rm min}$, by definition, so that $\phi_{\rm min}
\le \alpha_k \le 1$. $\Box$ 

\medskip
\noindent 
{\bf Approximation 1:} As a first heuristic approximation, we replace 
the full iteration expression of equation (\ref{iteralpha}) for 
$\alpha_k$ with the following simplified form 
\be
\alpha_{k+1} = {x \phi + x_k \over x + x_k}  \, , 
\label{firstalf} 
\ee 
i.e., we use $\alpha_k$ = 1 as an approximation for the previous value
[keep in mind that $x$ is the value at the ($k+1$)th cycle].  Using
equation (\ref{firstalf}) to evaluate $\alpha_k$ in the iteration
formula for ${\cal F}_k$, we obtain a working approximation for the
growth rate. Notice that $\alpha_k$ appears in the iteration formula
for ${\cal F}_k$, so that we must use equation (\ref{firstalf})
evaluated at $k$ rather than $k+1$.  As a result, the iteration factor
${\cal F}_k$ involves the random variables $x_k$ from three cycles,
or, equivalently (since the $x_k$ are i.i.d.) three separate samplings
of the variables.  We change notation so that $x_{j1}, x_{j2}, x_{j3}$
denote the three independent samplings of the random variables $x_k$.
Similarly, let $\phi_{j1}, \phi_{j2}$ denote two independent samplings
of the $\phi_k$. The iteration formula for this approximation can then
be written in the form
\be 
{\cal F}_j = 1 + { x_{j1}^2 \phi_{j1} (x_{j2} + x_{j3}) + x_{j2}
  (x_{j2} \phi_{j2} + x_{j3} ) \over x_{j1} \left[ (x_{j2} + x_{j3}) +
    x_{j2} (x_{j2} \phi_{j2} + x_{j3} ) \right] } \, . 
\label{iterapproxone} 
\ee
The growth rate for matrix multiplication can then be approximated by 
\be
\gamma = \lim_{N \to \infty} {1 \over N} \sum_{j=1}^N \log {\cal F}_j  \, , 
\label{approxgamma} 
\ee
where ${\cal F}_j$ is given by equation (\ref{iterapproxone}). As a
consistency check, for the restricted problem where the $\phi_{jn} = 1$,
the iteration factor ${\cal F}_j$ reduces to that appropriate for the
highly unstable regime (see equation [\ref{result4}]).

\medskip
\noindent 
{\bf Approximation 2:} To derive a second approximation for the growth
rate, we need a better approximation for the $\alpha_k$. If the values
of $x_k$ and $\phi_k$ were constant, then the $\alpha_k$ would
approach a constant value given by
\be
\alpha_k = {1 \over 2} \left\{ (1 - x_k/ x_{k-1}) + \left[ 
(1 - x_k/ x_{k-1})^2 + 4 (x_k/ x_{k-1}) \phi_k \right]^{1/2} \right\} \, . 
\label{zeropoint} 
\ee 
Even though the $x_k$ and $\phi_k$ are not constant, and the
$\alpha_k$ vary, we can use equation (\ref{zeropoint}) as an
approximation to specify the values of $\alpha_k$ appearing in the
exact formula of equation (\ref{gammafull}) for the growth rate.
After using this form to specify the $\alpha_k$, and relabeling the
indices, the iteration factor takes the form
\be
{\cal F}_k = 1 + 
{ x_{k1}^2 \phi_{k1} 2 x_{k3} + x_{k2} \left\{ (x_{k3} - x_{k2}) + \left[
(x_{k3} - x_{k2})^2 + 4 x_{k2} x_{k3} \phi_{k2} \right]^{1/2} \right\} 
\over
x_{k1} \left( 2 x_{k3} + x_{k2} \left\{ (x_{k3} - x_{k2}) + 
\left[ (x_{k3} - x_{k2})^2 + 4 x_{k2} x_{k3} \phi_{k2} \right]^{1/2} 
\right\} \right) } \, . 
\label{approxtwo} 
\ee
In the case $\phi_{jn} = 1$, the iteration factor of equation
(\ref{approxtwo}) reduces to the expression for the highly unstable
regime (Result 4).

\begin{figure} 
\centering
\includegraphics[width=120mm]{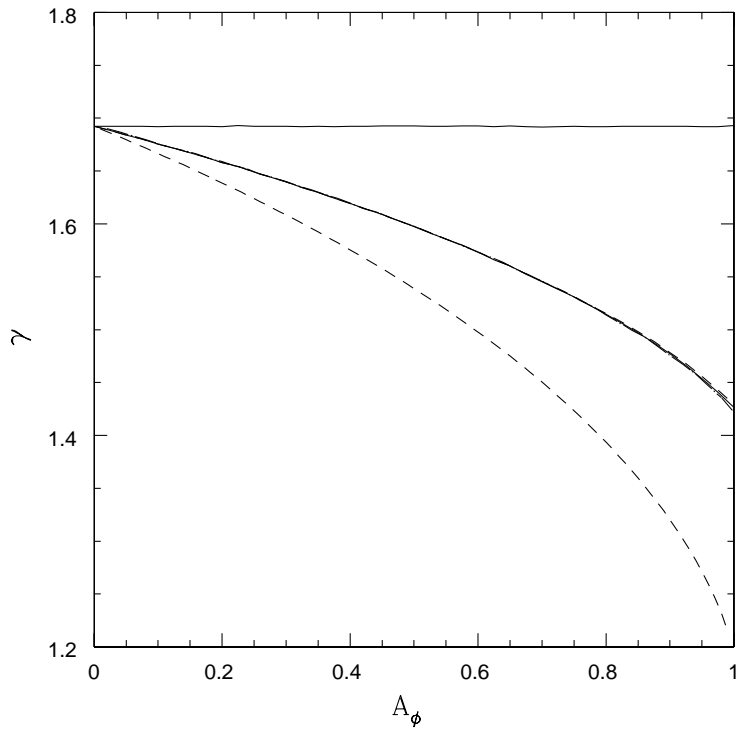}
\caption{Validity of approximations of equation (\ref{approxgamma}) 
and equation (\ref{approxtwo}) as a function of the deviation of
$\phi_k$ from unity. The upper solid line shows the growth rate for
matrix multiplication in the highly unstable regime where $\phi_k = 1$. 
The lower solid curve shows the growth rate for the case where 
$\phi_k = 1 - \amp \xi_k$, where $\xi_k$ is a uniformly distributed random
variable $0 \le \xi_k \le 1$. The dotted curve shows the estimate for
growth rate calculated from equation (\ref{approxgamma}) using the
same sampling of the $\phi_k$ variables; similarly, the dot-dashed curve
shows the approximation of equation (\ref{approxtwo}). Notice that both
of these approximations are almost identical to the actual result. The
dashed curve shows the lower limit to the growth rate derived in [AB]. } 
\label{fig:gamamp} 
\end{figure}

Figure \ref{fig:gamamp} shows how well these two approximation schemes
work.  The $\phi_k$ variables are chosen from the expression $\phi_k$ =
$1 - \amp \xi_k$, where $\xi_k$ is a random variable uniformly sampled
from the interval $0 \le \xi_k \le 1$ and where $\amp$ sets the
amplitude of the departures of the $\phi_k$ from unity. The growth
rate is shown as a function of the amplitude.

In [AB], we derived a bound on the difference between the growth rate
for the general case $\gamma$ (considered here) and the growth rate in the
highly unstable regime $\gamma_0$, i.e., 
\be 
\gamma_0 - \gamma \le {1 \over 2} \langle \log \phi_k \rangle \, . 
\ee 
This bound is shown as the dashed curve in Figure \ref{fig:gamamp}.
The true growth rates fall comfortably between this lower bound and
the growth rate for the highly unstable regime (where the latter 
provides an upper bound). 

Thus far, this paper has focused on the regime where the transformation
matrices are classically unstable. Before considering classically
stable matrix multiplication in the next section, we note the
following result that applies at the transition between the two
regimes:

\medskip 
\noindent 
{\bf Result 6:} Consider the matrix transformation that maps the
principal solutions from one cycle to the next.  When the matrix
elements $g_k = {\dot y}_1 (\pi)$ vanish, then the remaining matrix
elements are $h_k = y_1 (\pi) = \pm 1$. The transformation matrix 
${\mfont M}_{g0}$ for this case is stable under multiplication.
 
(The proof is a simple explicit computation.)  

\section{Elliptical Rotations and the Classically Stable Regime} 
 
When the principal solutions $h_k$ appearing in the discrete map of
equation (\ref{mapzero}) are less than unity, matrix multiplication is
stable for the case of constant parameters. In the case of interest,
however, the parameters in Hill's equation (\ref{basic}) and the
matrices (\ref{mapzero}) vary from cycle to cycle. This section
considers the case where the $|h_k| \le 1$, but vary from cycle to
cycle, and show that instability results.  In this regime, the
discrete map takes the form of an elliptical rotation matrix [LR] as
described below.  We thus find the growth rates for elliptical
rotation matrices for the case where the matrix elements vary from
cycle to cycle.

\medskip  
\noindent
{\bf Definition:} An {\it elliptical rotation matrix} is defined to be 
\be
{\mfont E} (\theta; L) \equiv \left[ 
\matrix{ \cos\theta & - L \sin\theta \cr 
(1/L) \sin\theta & \cos \theta }  
\right] \, . 
\label{ellipdef} 
\ee
 
\noindent
These matrices have the following properties: 
 
\noindent
The product of elliptical rotation matrices with the same value of $L$ 
produces another elliptical rotation matrix, also with the same $L$, 
\be
{\mfont E} (\theta_1; L) {\mfont E} (\theta_2; L) = 
{\mfont E} \left( [\theta_1 + \theta_2]; L \right) \, . 
\ee
As a result, the elliptical rotation matrices form a group. 

\noindent
For fixed $L$, matrix multiplication is stable. Specifically, the
eigenvalues of the product of $N$ matrices (with fixed $L$) have the
form
\be
\lambda = \exp \left[ \pm i \sum_{j=1}^N \theta_j \right] \, = 
\exp \left[ \pm i \theta_N \right] \, , 
\ee
where $\theta_N$ is the angle corresponding to the group element 
produced after $N$ matrix multiplications.  

\medskip 
\noindent
{\bf Result 7:} When an individual cycle of Hill's equation is
stable, specifically when $|h_k| \le 1$, the full transformation
matrix ${\mfont M}_k$ takes the form of an elliptical rotation.

\noindent
{\it Proof:} Since $|h_k| \le 1$, we can define an angle $\theta_k$ such
that $h_k = \cos \theta_k$. The full matrix ${\mfont M}_k$ given by 
equation (\ref{mbdefine}) then takes the form
\be 
{\mfont M}_k = \left[ \matrix{ \cos \theta_k & - (\sin^2 \theta_k)/g_k \cr 
g_k & \cos \theta_k } \right] \, = 
\left[ \matrix{ \cos \theta_k & - L_k \sin \theta_k \cr 
(1/L_k) \sin \theta_k & \cos \theta_k } \right] \, = {\mfont E}_k (\theta_k; L_k) \, , 
\ee
where we have defined $L_k = (\sin\theta_k)/g_k$. As before, we can
factor out the $\cos\theta_k = h_k$ and write the matrix in the form
\be 
{\mfont M}_k = 
\cos\theta_k \left[ \matrix{1 & x_k \phi_k \cr 1/x_k & 1} \right] \, = \, 
\cos\theta_k {\mfont B}_k \, , 
\label{anotherbdef}
\ee
where 
\be
x_k = L_k / \tan\theta_k \qquad {\rm and} \qquad \phi_k = - \tan^2\theta_k \, . 
\label{transellipse} 
\ee
Equation (\ref{transellipse}) thus specifies the transformation
between the random variables $(x_k, \phi_k)$ appearing in the original
transformation matrix and the random variables $(\theta_k, L_k)$ in
the corresponding elliptical rotation matrix.  Note that the values of
$\phi_k$ are strictly negative in this formulation. Otherwise, the
matrix ${\mfont B}_k$ has the same form as in equation
(\ref{mbdefine}). $\Box$

If we let $\gamma_B$ be the growth rate for matrix ${\mfont B}_k$,
then the growth rate $\gamma_M$ for the full matrix ${\mfont M}_k$
takes the form
\be
\gamma_M = \gamma_B + \lim_{N \to \infty} {1 \over N} \sum_{k=1}^N 
\log [ \cos \theta_k ] \, . 
\label{gammasum} 
\ee
The exact growth rate for the matrix ${\mfont B}_k$ (see equation
[\ref{anotherbdef}]) is given by Theorem 1. In particular, equations
(\ref{gammafull}) and (\ref{iteralpha}) remain valid for negative
values of the $\phi_k$ and can be used to calculate the growth rate.

\medskip
\noindent
{\bf Result 8:} For an elliptical rotation matrix with constant
angle $\theta$ and random $L_k$, the growth rate for matrix
multiplication vanishes in the two limits $h = \cos\theta \to 0$ and 
$h = \cos\theta \to 1$. 

\noindent 
{\it Proof:} In the limit $h \to 1$ we have  $\sin\theta$ = 0, and the
elliptical rotation matrix becomes the identity matrix. As a result,
the growth rate vanishes. 

\noindent
In the other case where $h \to 0$, $\sin\theta$ = 1, and the matrix
takes the form
\be
{\mfont E}_k \to {\mfont E}_{0k} = 
\left[ \matrix{0 & -L_k \cr 1/L_k & 0 } \right] \, . 
\ee
In this case, for even numbers of matrix multiplications, 
say $N$ = $2n$, the product matrix takes the form 
\be
{\mfont E}_{0k}^{(N)} = \prod_{k=1}^N {\mfont E}_{0k} = 
(-1)^n \left[ \matrix{P^A_n & 0 \cr 0 & P^B_n } \right] \, , 
\ee
where the matrix elements are given by the products 
\be
P^A_n = \prod_{k=1}^{n} {L_{2k} \over L_{2k-1}} 
\qquad {\rm and} \qquad 
P^B_n = \prod_{k=1}^{n} {L_{2k-1} \over L_{2k}} \, . 
\label{pndef} 
\ee
The eigenvalues of the product matrix are given by $\lambda = P^A_n$
and $\lambda = P^B_n$. For odd $N = 2n+1$, the eigenvalue $|\lambda|$
= $(P^A_n P^B_n)^{1/2}$.  In either case, in the limit of large $N$,
the growth rate for matrix multiplication takes the form
\be
\gamma = \lim_{N \to \infty} {1 \over N} \sum_{k=1}^N \log 
\left[ {L_{2k} \over L_{2k-1} } \right] = 
\left\langle \log L_{2k} \right\rangle - 
\left\langle \log L_{2k-1} \right\rangle \, = 0 \, .
\ee
The final equality holds because the $L_k$ are independent. $\Box$

Elliptical rotation matrices are unstable under multiplication when
their parameters vary from cycle to cycle: 

\medskip
\noindent 
{\bf Theorem 4:} Consider an elliptical rotation matrix with variable
angle $\theta_k$ and symmetric fluctuations of the $L_k$ parameter
about its mean value $L_0$. The variations are thus written in the
form $L_k$ = $L_0 (1 + \eta_k)$, where the odd moments 
$\langle \eta_k^{2n+1} \rangle$ = 0 for all integers $n$.  For small
fluctuations $|\eta_k| < 1$, the growth $\gamma$ rate for matrix
multiplication takes the form
\be
\gamma = {1 \over 2} \lim_{N\to\infty} {1 \over N} 
\sum_{k=1}^N \log \left[ \cos^2 \theta_k + \sin^2 \theta_k 
\left\langle {1 \over 1 + \eta_{j}} \right\rangle \right] + 
{\cal O} \left( \eta_k^4 \right) \, .  
\label{quadellipse} 
\ee

\noindent 
{\it Proof:} We first break up the matrix into two parts so that 
\be
{\mfont E}_k = {\mfont I} \cos \theta_k + \sin \theta_k {\mfont Z}_k \, , 
\ee 
where $\mfont{I}$ is the identity matrix and where 
\be
{\mfont Z}_k = \left[ \matrix{0 & -L_k \cr 1/L_k & 0} \right] \, . 
\ee
The product of $N$ matrices $\mfont{E}_k$ becomes 
\be
\mfont{E}^{(N)} = \prod_{k=1}^N \mfont{E}_k = 
\sum_{\ell=0}^N  \sum_{k=1}^{C_\ell^N} 
\left( \prod_{i=1}^{N-\ell} \cos \theta_i \right)_k  
\left( \prod_{j=1}^{\ell} {\mfont Z}_j \sin\theta_j \right)_k \, ,  
\ee
where the subscripts on the products denote different realizations. 
The products of even numbers $\ell = 2 n$ of matrices $\mfont{Z}_k$ 
produce diagonal matrices of the form
\be
{\mfont Z}^{(\ell)} = {\mfont Z}^{(2 n)} = 
\prod_{k=1}^n {\mfont Z}_{2k} {\mfont Z}_{2k-1} 
= (-1)^n \left[ \matrix{P_n^A & 0 \cr 0 & P_n^B} \right] \, , 
\ee
where the matrix elements $P^A_n$ and $P^B_n$ are given by equation
(\ref{pndef}).  Similarly, the product of odd numbers $\ell = 2n + 1$
of matrices $\mfont{Z}_k$ produce off-diagonal matrices of the form 
\be
{\mfont Z}^{(\ell)} = {\mfont Z}^{(2n+1)} = 
\left\{ \prod_{k=1}^n {\mfont Z}_{2k+1} {\mfont Z}_{2k} \right\} 
{\mfont Z}_1 = (-1)^n 
\left[ \matrix{0 & - P_n^A L_1 \cr P_n^B / L_1 & 0 } \right] \, , 
\ee
where the $P_n$ are defined previously. Next we write the expectation
values of these products in the form
\be
\left\langle P_n \right\rangle = \left\langle 
\prod_{j=1}^n {L_{2j} \over L_{2j-1}} \right\rangle = \left\langle 
\prod_{j=1}^n {1 + \eta_{2j} \over 1 + \eta_{2j-1}} \right\rangle = 
\left\langle {1 \over 1 + \eta_{j}} \right\rangle^n 
\equiv \rat^n \, . 
\label{pbar} 
\ee
This expression holds because the odd powers of the $\eta_j$ vanish in
the mean, and the samples of the different $\eta$'s are independent.

The eigenvalue $\Lambda_N$ of the product matrix at the $Nth$
iteration can be written in terms of its matrix elements, i.e.,
\be
\Lambda_N = \sigma_{11} + \sigma_{22} \, . 
\ee
Without loss of generality, let $N = 2 K$ be even. The matrix
elements $\sigma_{11} = \sigma_{22} = \sigma$ are given by
\be
\sigma = \sum_{m=0}^{K} \, \sum_{k=1}^{C_{2m}^{2K}} \, 
\left( \prod_{i=1}^{2K-2m} \cos \theta_i \right)_k 
\left( \prod_{i=1}^{2m} \sin \theta_i \right)_k  (-1)^m \, 
\rat^m \, , 
\ee
where $C^{2K}_{2m}$ is the binomial coefficient and where we have 
used equation (\ref{pbar}). This expression for $\sigma$ contains 
the even terms of a binomial expansion. We can thus write the 
eigenvalue in the form 
\be
\Lambda_N = \prod_{k=1}^N \left[ \cos \theta_k + i \sin \theta_k 
\rat^{1/2} \right]_k
+ \prod_{k=1}^N \left[ \cos \theta_k - i \sin \theta_k 
\rat^{1/2} \right]_k \, . 
\ee
Next we define 
\be
A_k \equiv \left[ \cos^2 \theta_k + \sin^2 \theta_k \rat \right]^{1/2} 
\qquad {\rm and} \qquad \tan \alpha_k \equiv \rat^{-1/2} \tan \theta_k \, . 
\ee 
The eigenvalue takes the form 
\be
\Lambda_N = 2 \, \left( \prod_{k=1}^N A_k \right) \, 
\cos \left( \sum_{k=1}^N \alpha_k \right) \, , 
\ee
and the corresponding growth rate becomes 
\be
\gamma = {1 \over 2} \lim_{N\to\infty} {1 \over N} 
\sum_{k=1}^N \log \left[ \cos^2 \theta_k + 
\sin^2 \theta_k \rat \right] \, . 
\label{egrow} 
\ee
Using the definition of $\rat$, we obtain the result of Theorem 4.
The order of the error term follows by comparing equation
(\ref{egrow}) with the leading order expansion [AB2].
$\Box$

In the regime of small $\eta_k \ll 1$, the expression for the growth 
rate reduces to the form 
\be
\gamma = {1 \over 2} \left\langle \sin^2 \theta_k \right\rangle
\left\langle \eta_k^2 \right\rangle \, .  
\ee

This section shows that instability does not require a finite
threshold for the amplitude of the fluctuations in $L_k$.  Nonzero
amplitude leads to instability with growth rate $\gamma \propto
\langle \eta_k^2 \rangle$. Variations in the original parameters
$(\lambda_k, q_k)$ of Hill's equation lead to fluctuations in the
principal solutions $(h_k, g_k)$; fluctuations in the $(h_k, g_k)$
lead to variations in the $L_k$ and hence growth. As a result, Hill's
equation with random forcing terms is generically unstable. One
notable exception occurs when the $h_k$ = 0 or $h_k$ = 1 (Result 8).
 
\section{Conclusion}

This paper provides expressions for the growth rates for the random 
$2 \times 2$ matrices that result from solutions to the random Hill's 
equation (\ref{basic}). Theorem 1 gives an exact expression for the
growth rate. Theorems 2 and 3 provide approximate growth rates for the
regimes where the variables $\phi_k$ are small, and close to unity,
respectively. Additional approximations for are given in Section 4.
When Hill's equation is classically stable, the discrete map that
governs the solutions has the form of an elliptical rotation matrix
(equ. [\ref{ellipdef}]).  With fixed elements, such matrices are
stable under multiplication; variations in the $L_k$ parameter lead to
instability.  For small symmetric fluctuations of the length parameter
$L_k$, the growth rate is given by Theorem 4.  

\medskip

\begin{acknowledgements}
We would like to thank Scott Watson and Michael Weinstein for useful
conversations and suggestions. The work of FCA and AMB is jointly
supported by NSF Grant DMS-0806756 from the Division of Applied
Mathematics, and by the University of Michigan through the Michigan
Center for Theoretical Physics. AMB is also supported by the NSF
through grants DMS-0604307 and DMS-0907949. FCA is also supported by
NASA through the Origins of Solar Systems Program via grant
NNX07AP17G.
\end{acknowledgements}


\begin{thebibliography} {}

\bibitem[AS]{abstegun}
M. Abramowitz and I. A. Stegun, 
{\it Handbook of Mathematical Functions}, Dover, New York, 1970.

\bibitem[AB]{adamsbloch}  
F. C. Adams and A. M. Bloch, 
{\it Hill's Equation with random forcing terms}, 
SIAM J. Appl. Math., 68 (2008), pp. 947 -- 980.  

\bibitem[AB2]{adamsbloch2}  
F. C. Adams and A. M. Bloch, 
{\it Hill's equation with random forcing parameters:} 
{\it The limit of delta function barriers}, 
J. Math. Phys., 50 (2009), pp. 073501, 1- 20.  

\bibitem[AK]{adamsapj} 
F. C. Adams, A. M. Bloch, S. C. Butler, J. M. Druce, and J. A. Ketchum, 
{\it Orbits and instabilities in a triaxial cusp potential}, Astrophys. J., 
670 (2007), pp. 1027 -- 1047. 

\bibitem[AN]{anderson} 
P. W. Anderson, {\it Absence of diffusion in certain random lattices},
Physical Review, 109 (1958), pp. 1492 -- 1505. 

\bibitem[CL]{carmonalacroix}
R. Carmona and J. Lacroix
{\it Spectral Theory of Random Schroedinger Operators}
Birkhauser, Boston, 1990.

\bibitem[CR]{cambronero} 
S. Cambronero, B. Rider, and J. Ram{\'i}rez, {\it On the shape of}
{\it the ground state eigenvalue density of a random Hill's equation}, 
Comm. Pure Appl. Math., 59 (2006), pp. 935 -- 976.   

\bibitem[CN]{cohennewman} 
J. E. Cohen and C. M. Newman, {\it The stability of large random}
{\it matrices and their products}, Annals of Prob., 12 (1984), pp. 283 -- 310. 

\bibitem[FU]{furst} 
H. Furstenberg, {\it Noncommuting random products}, Trans. Amer. 
Math. Soc., 108 (1963), pp. 377 -- 428. 

\bibitem[FK]{furstkest} 
H. Furstenberg and H. Kesten, {\it Products of random matrices}, 
Ann. Math. Statist., 31 (1960), pp. 457 -- 469. 

\bibitem[HI]{hill} 
G. W. Hill, {\it On the part of the motion of the lunar perigee}
{\it which is a function of the mean motions of the Sun and Moon},
Acta. Math., 8 (1886), pp. 1 -- 36. 

\bibitem[LGP]{lifshitz}
I. Lifshitz, S. Gredeskul, and L. Pastur, 
{\it Introduction in the Theory of Disordered Systems}, Wiley, New York, 1988.

\bibitem[LR]{limarahibe}
R. Lima and M. Rahibe, {\it Exact Lyapunov exponent for infinite products} 
{\it of random matrices}, J. Phys. A. Math. Gen. 27 (1994), pp. 3427 -- 3437. 

\bibitem[MW]{magwinkler}
W. Magnus and S. Winkler, {\it Hill's Equation}, Wiley, New York, 1966. 

\bibitem[O]{Oseledec}
V.I. Oseledec, {\it A multiplicative ergodic theorem, Lyapunov
charateristic numbers for dynamical systems},
Trans. Moscow Mathematcical Society, 19 (1968), 197-231.

\bibitem[PF]{pasturfigotin}
Pastur, L., and Figotin, A., 
{\it Spectra of Random and Almost-Periodic Operators}, a Series of
Comprehensive Studies in Mathematics, (Springer-Verlag, Berlin, 1991).

\bibitem[PI]{pincus} 
S. Pincus, {\it Strong laws of large numbers for products of random matrices},
Trans. Amer. Math. Soc., 287 (1985), pp. 65 -- 89. 

\end{thebibliography}
\end{document}